\documentclass[conference]{IEEEtran}
\usepackage{blindtext}
\usepackage{graphicx}
\usepackage{setspace}
\usepackage{bm, amsmath, amssymb,amsthm}
\usepackage{lipsum,multicol}
\usepackage{cite}
\usepackage{epsfig}
\usepackage{bigints}
\usepackage{cases}
\usepackage{subcaption}
\usepackage{color}
\usepackage{graphics}
\usepackage{amsfonts}
\usepackage{etoolbox}
\usepackage{float}
\usepackage{soul}
\usepackage{caption,amsmath,amssymb,epstopdf,array,amsfonts,float,multicol,cite,xcolor,subcaption,caption,multicol,multirow}
\usepackage[font=small]{caption}
\usepackage{tikz}
\soulregister\cite7
\soulregister\ref7
\soulregister\pageref7
\soulregister\eqref7
\DeclareGraphicsExtensions{.pdf,.png,.jpg}
\graphicspath{{Figures/}}
\DeclareMathOperator{\E}{\mathbb{E}}
\DeclareMathOperator{\C}{\mathbb{C}}

\newcommand{\hsr}{|h_{sr}|^2}
\newcommand{\hrd}{|h_{rd}|^2}
\newcommand{\hsd}{|h_{sd}|^2}
\newcommand{\gsp}{|g_{sp}|^2}
\newcommand{\grp}{|g_{rp}|^2}
\newcommand{\lsr}{\lambda_{sr}}
\newcommand{\lsd}{\lambda_{sd}}
\newcommand{\lrd}{\lambda_{rd}}
\newcommand{\lsp}{\lambda_{sp}}
\newcommand{\lrp}{\lambda_{rp}}
\newcommand{\gth}{\gamma_{th}}
\newcommand\copyrighttext{ %
	\footnotesize ``This work has been submitted to IEEE  WPMC 2017 conference for possible publication. Copyright may be transferred without prior notice, after which this version may no longer be accessible"}
\newcommand\copyrightnotice{%
	\begin{tikzpicture}[remember picture,overlay]
	\node[anchor=south,yshift=760pt] at (current page.south) {{\parbox{\dimexpr\textwidth-\fboxsep-\fboxrule\relax}{\copyrighttext}}};
	\end{tikzpicture}%
}
\begin{document}
\title{Performance of DF Incremental Relaying with Energy Harvesting Relays in Underlay CRNs} 
\author{
	\IEEEauthorblockN{Komal Janghel and Shankar Prakriya\\
		Department of Electrical Engineering,\\
		Indian Institute of Technology, Delhi, New Delhi, India\\
		E-mail: komal.janghel@gmail.com, shankar@ee.iitd.ac.in}
\thanks{\hspace*{-0.2cm}\hrulefill}
\thanks{This work was supported by Information Technology Research Academy (ITRA) through sponsored project ITRA/15(63)/Mobile/MBSSCRN/01.}
}
\maketitle
\begin{abstract}
In this paper, we analyze the throughput performance of  incremental relaying using energy harvesting (EH) decode-and-forward (DF) relays in underlay cognitive radio networks (CRNs). The destination combines the direct and relayed signals when the direct link is in outage. From the derived closed-form expressions, we present an expression for the power-splitting parameter of the EH relay that optimizes the throughput performance.  We demonstrate that relaying using EH DF relays results in better performance than direct signalling without a relay {\em only} when the destination combines the direct signal from the source with the relayed signal. Computer simulations demonstrate accuracy of the derived expressions. 
\end{abstract}
\copyrightnotice
\IEEEpeerreviewmaketitle
\section{Introduction}
Cognitive radio networks (CRNs) have shown great promise in alleviating the acute shortage of spectrum. In such networks, secondary (unlicensed) users are allowed to share the spectrum of the primary (licensed) users. Underlay  CRNs in particular, have been shown to result in great improvement in spectral utilization efficiency. In these networks, the  secondary transmitters transmit simultaneously with the primary transmitters in the same frequency band, but with powers carefully constrained to limit interference to the primary receiver below a specified interference temperature limit. 
\par In the recent years, the use of energy harvesting (EH)  is being studied to prolong battery life of nodes. Although energy can be harvested from natural sources, it is wireless EH that shows the greatest  promise. In particular, the possibility of simultaneous wireless  information and power transfer has spurred research interest in this area. Since practical circuits cannot simultaneously harvest energy and perform information processing, time-switching and power-splitting relaying protocols have been proposed \cite{Nasir2013}. In the former, the relay is first charged by the source for a fraction of the signalling interval prior to two-hop relaying. In the latter, the received signal is split into two parts, with a fraction (referred to as the power-splitting parameter $\rho$) being used for energy harvesting, while the rest is used for information processing. It is well known that optimization of this parameter is crucial to maximize throughput. 
\par While the use of EH relays in CRNs is well motivated, analysis of performance of such networks has attracted attention only over the past few years 
  \cite{Janghel2014,Liu2015,Bhowmick2016,Yang2015,Miridakis2016,Wang2016,He2016,Janghel2016}. 
 Two types of EH methodologies have been proposed in the context of CRNs In the first type, energy is harvested from the primary signal \cite{Janghel2014,Liu2015,Bhowmick2016,Miridakis2016,
 Wang2016,He2016}. Note that \cite{Bhowmick2016,Miridakis2016} use the interweave cognitive radio principles, \cite{Janghel2014,Liu2015,Janghel2016} use the underlay signalling mode, while \cite{Wang2016,He2016} use overlay principles.  In the second type, which is of primary interest in this paper, energy is harvested from the secondary source \cite{Yang2015,Janghel2016} (this requires secondary EH nodes to be in close proximity to the secondary transmitter). Under these circumstances, even in the non-cognitive context, the importance of considering the direct channel from source to destination in addition to the signal relayed by the EH relay has been recognized by only a few authors \cite{Lee2017,Van2016}. In {\em all existing literature} on two-hop EH relaying in underlay CRNs \cite{Liu2015,Yang2015,Janghel2016}, the direct channel from source to destination has been ignored. In underlay cognitive radio, powers used at secondary transmitters is a random quantity. For this reason, the secondary nodes need to be in close proximity to each other to ensure any reasonable quality of service in the secondary links. In underlay signalling with EH relays, the nodes need to be even closer since energy harvested is typically small. Ignoring the direct channel from source to destination is therefore not reasonable in such situations. In this paper, we demonstrate this fact. 
\par The contributions of this paper are as follows:
\begin{enumerate}
	\item We derive closed-form expressions for the throughput of an incremental relaying scheme in which the destination combines the  signal relayed by the EH relay with the direct signal from the source. 
	\item Using approximated throughput expressions, we derive an expression for the power-splitting parameter that maximizes throughput. 
	\item We demonstrate that relaying with an EH relay results in larger throughput than direct signalling only when the destination combines the direct and relayed signals. 
\end{enumerate}
{\em Notations}: $\E_{\C}(\cdot)$ denotes the expectation over the condition/conditions $\C$. $\exp(\lambda)$ represents the exponential distribution with parameter $\lambda$, and ${\cal CN}(0,a)$ denotes the circular normal distribution with mean $0$ and variance $a$. $E_1(\cdot)$ and $\text{Ei}(\cdot)$ represent the exponential integrals defined in \cite[5.1.1]{Abramowitz1964} and \cite[5.1.2]{Abramowitz1964} respectively. 
\section{System Model}\label{sec:system_model}
\par
\begin{figure}[h]
	\vspace*{-0.3cm}
\centering
\includegraphics[width=0.5\textwidth]{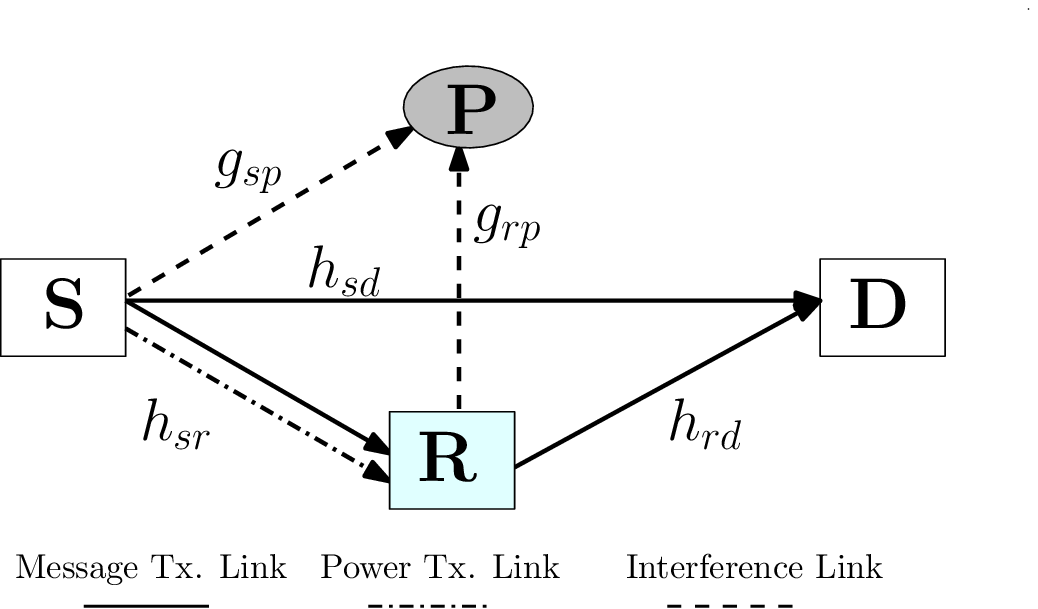}
\caption{System Model}
\label{fig:sys_model}
\vspace*{-0.3cm}
\end{figure}

 We consider a two-hop decode-and-forward (DF) relay network as depicted in Fig.~\ref{fig:sys_model}. 
The primary network consists of the primary transmitter (not depicted in the figure) and the primary receiver P. The secondary network (SN) consists of three nodes -  a source (S), a relay (R) and a destination (D). Each node equipped with one antenna.  Denote the channel between S and R by $h_{sr}\sim {\cal CN}(0,\lambda_{sr}^{-1})$,  and that between the R and D by $h_{rd}\sim {\cal CN}(0,\lambda_{rd}^{-1})$. Similarly,  denote the channel between S and P by   $g_{sp}\sim {\cal CN}(0,\lambda_{sp}^{-1})$, and that between R and P by $g_{rp}\sim {\cal CN}(0,\lambda_{rp}^{-1})$. We assume  that R is equipped with a super-capacitor, and acts as an EH node with EH factor $\eta$.  The energy harvested in the first phase is used by R to relay the signal to D.  As in most literature on underlay CRN, we neglect  the primary signal at R and D. This is reasonable because  of  the large distance between the primary transmitter and the secondary nodes  \cite{Tourki2013} (this assumption has been justified on information theoretic grounds \cite{Jovicic2009}). All the channels are reversible and quasi-static.
\section{Transmission Protocol}\label{sec:transmission_protocol} 
\begin{figure}[h]
	\centering
	\includegraphics[width=0.48\textwidth]{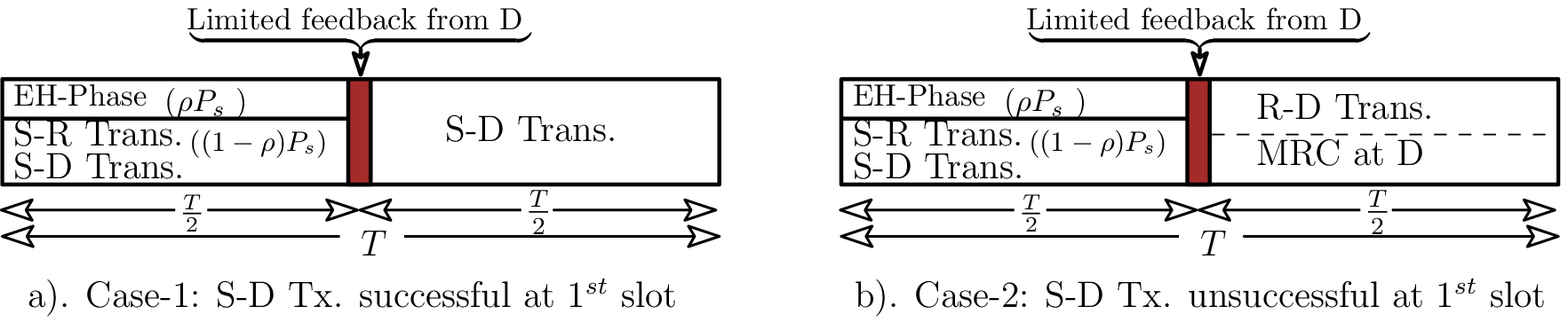}
	\caption{Transmission scheme for incremental relaying}
	\label{fig:transmission_schemeInc}
\end{figure}

We assume fixed-rate transmission at rate $R_s$ by all nodes. Signalling based on the incremental protocol is completed in two time-slots as depicted in Fig.\ref{fig:transmission_schemeInc}. Message transmission is based on the incremental relaying protocol \cite{Ikki2011}, and EH is based on the power-splitting  protocol \cite{Nasir2013}.
\par In the first transmission time-slot, S transmits information symbols to D and R as depicted in Fig.\ref{fig:transmission_schemeInc}. R harvests energy from this signal using power splitting, and attempts to decode the information symbols. Meanwhile,  D attempts to decode the symbols. If successful, it sends feedback to the source and the relay\footnote{We assume that feedback time is extremely small and can be neglected in the analysis without loss of generality.}. The relay discards the decoded symbols, and the source re-transmits a new block of symbols to D as depicted in Fig.\ref{fig:transmission_schemeInc}a\footnote{The relay does not harvest energy in this phase. Energy stored in the super-capacitor is assumed to be lost since it lacks the ability to store charge over long intervals.}. Else, R relays the symbols using the energy harvested, and D combines the signals in the first and the second time-slots as depicted  in Fig.\ref{fig:transmission_schemeInc}b.
\par In the first time-slot, S transmits a message signal $x$ with power $P_s$ in underlay mode to R and D. In PS-EH case, a component $y_{d_1}$ of the received signal with $\rho$ fraction of the power is utilized for EH, while the remaining signal with fraction $1-\rho$ of the power is used to decode the symbols. Clearly, $y_r$ at R and and  $y_{d_1}$ at D in the first phase are given by:
\begin{IEEEeqnarray}{rcl}
y_r &=& \sqrt{(1-\rho) P_{s}}\,x\,\hsr+n_{r}\;\quad\text{and}\label{eq:yr_PS}\\
y_{d_1} &= &\sqrt{P_{s}}\,x\, \hsd+n_{d_{1}} \label{eq:yd1_PS}
\end{IEEEeqnarray}
respectively, where $n_r,n_{d_1}\sim{\cal CN}\left(0,N_o\right)$ are  noise samples at R and D. Clearly, the SNR $\Gamma_{d_1}$ at D in the first time-slot is $\frac{P_s\hsd}{N_o}$. Energy harvested at R is $\eta \rho\,P_s \hsr/2$ (ignoring noise) so that the  power available for relaying is given by:
\begin{equation}
^hP_{r}=\rho\,\eta P_s \hsr = \beta\,P_s \hsr\,,\label{eq:Prh-PS}
\end{equation}
where $\beta= \eta\, \rho$.
 Let $I$ denote the interference temperature limit.   In order to ensure that the interference caused to primary receiver P is limited to $I$, the power $P_s$ at S  is chosen to be:
\begin{equation}
P_s= {I}/{\gsp}\,.\label{eq:PsOnlyInterference}
\end{equation}
 We consider only the peak interference constraint at S, and ignore the peak power constraint for the following reason:
 \begin{enumerate}
 \item It is well known that performance of CRNs exhibits an outage floor, and does not improve with increase in peak power (it is in this low outage and high throughput region that CRNs are typically operated) \cite{Lee2011,Duong2012} (and references therein). In this paper, we discuss optimization of PS EH parameter, which is of interest in this high throughput regime.
 \item Since performance of CRNs is typically limited by interference, and sufficient peak power is typically available, this assumption is quite reasonable.
 \end{enumerate} 
\par In the second time-slot, R is used to forward the decoded symbol $\hat{x}$ to D. In order to ensure that the interference at P is constrained to $I$, the total transmit power $ P_{r} $ at R is chosen to be:
\begin{equation}
 P_{r} = \min\left(^hP_{r}, \frac{I}{|g_{rp}|^2} \right). \label{eq:Pr_PS}
\end{equation}
Received signal $y_{d_2}$ at D can be expressed as
\begin{equation}
y_{d_2} = \sqrt{P_{r}}\, \hat{x}\,h_{rd}+n_{d_2}\;, \label{eq:yd2_PS}
\end{equation} 
where $n_{d_2}\sim{\cal CN}\left(0,N_o\right)$ is the additive white Gaussian noise sample.  We assume that D uses MRC to combine the signals obtained in the first phase (\ref{eq:yd1_PS}) and second phase (\ref{eq:yd2_PS}). 
 Signal-to-noise ratios (SNRs) at R and D  are expressed using (\ref{eq:yr_PS}), (\ref{eq:yd1_PS}) and (\ref{eq:yd2_PS}) as:
\begin{IEEEeqnarray}{rcl}
	\Gamma_{r} &=&\frac{(1-\rho)P_s\hsr}{N_o}\qquad \text{and} \label{eq:SNR_rps}\\
	\Gamma_{d} &=& \underbrace{\frac{P_s\hsd}{N_o}}_{\Gamma_{d_1}}+\underbrace{\frac{P_{r}|h_{rd}|^2}{N_o}}_{\Gamma_{d_2}}.\label{eq:SNR_dps}
\end{IEEEeqnarray}
\section{Performance Analysis}\label{sec:perfromance}
In this section, we analyze throughput performance of the described incremental relaying protocol. When $\Gamma_{d_1}\geq\gamma_{th}$, the direct link is successful, so that rate achieved is $R_s$. When $\Gamma_{d_1}<\gamma_{th}$,  and the relay can decode successfully ($\Gamma_r\geq\gamma_{th}$), and the SNR at the destination after combining is sufficient ($\Gamma_d=\Gamma_{d_1}+\Gamma_{d_2}\geq \gamma_{th}$), signalling is completed in two hops (rate $R_s/2$). Clearly, throughput $\tau$ is given by:  
{\begin{IEEEeqnarray}{rcl}
		\tau&=& 0.5\,R_s\underbrace{\Pr\left(\Gamma_{d_1} < \gamma_{th},\, \Gamma_{r}\geq\gamma_{th},\, \Gamma_{d_1}+\Gamma_{d_2}\geq\gamma_{th}\right)}_{q_1}\nonumber\\
		&&\hspace{1.5in}+\underbrace{\ R_s\Pr(\Gamma_{d_1}\geq\gamma_{th})}_{q_2}\,,\label{eq:tauIn}
\end{IEEEeqnarray}}
where $\gth=2^{R_s}-1$. We note that $q_1$, the probability that the relayed link is successful while the direct link is not successful, can alternatively be represented as:
\begin{IEEEeqnarray}{rcl}
	q_1&=&1-\Big(\underbrace{\Pr\left(\Gamma_r< \gamma_{th}\right)}_{p_1}+\underbrace{\Pr\left(\Gamma_{d_1}\geq\gamma_{th},\Gamma_r \geq\gamma_{th} \right)}_{p_2}\nonumber\\
	&&\hspace{0.8in}+\underbrace{\Pr\left(\Gamma_d< \gamma_{th},\;\Gamma_r \geq \gamma_{th}\right)}_{p_3}\Big).\label{eq:q}
\end{IEEEeqnarray}
We evaluate each of the terms in what follows. From \eqref{eq:SNR_rps}, $p_1$ can be evaluated as:
\begin{eqnarray}
p_1=\Pr\left(\Gamma_r< \gamma_{th}\right) = \Pr\left(\frac{(1-\rho) P_s\hsr}{N_o}< \gamma_{th}\right). \label{eq:p1}
\end{eqnarray}
Using $P_s=I/|g_{sp}|^{2}$, it can be shown that: 
\begin{equation}
p_1=1-\frac{1}{1+{\lsr\psi}/({\lsp(1-\rho)})},\label{eq:p1Final}
\end{equation}
where $\psi=\frac{\gth}{I/N_o}$.
Similarly,  $q_2$ is given by: 
\begin{eqnarray}
q_2=\Pr(\Gamma_{d_1}\geq\gamma_{th})=\frac{1}{1+{\lsd \psi}/{\lsp}}.\label{eq:q2}
\end{eqnarray}
We note that $\Gamma_r$ and $\Gamma_{d_1}$ are not independent due to their dependence on the random variable $P_s=I/|g_{sp}|^{2}$. By first conditioning on $|g_{sp}|^{2}$, exploiting the independence of $\Gamma_{r\big|\gsp}$ and $\Gamma_{d\big||g_{sp}|^{2}}$, and then averaging over $|g_{sp}|^{2}$, we can show that  $p_2$ can be written as:
\begin{IEEEeqnarray*}{rcl}
p_2& =& \int_{0}^{\infty} \lsp e^{\lsp\gsp} \int_{\psi \gsp}^{\infty} \lsd e^{-\lsd \hsd} \int_{\frac{\psi\gsp}{(1-\rho)}}^{\infty}\lsr \nonumber\\
&&\hspace*{2cm}e^{-\lsr\hsr} d\hsr d\hsd d\gsp. \nonumber
\end{IEEEeqnarray*}
After integrating over $\hsr$ and $\hsd$, $p_2$ can be expressed in terms of $|g_{sp}|^{2}$ as: 
\begin{equation*}
p_2=\int_{0}^{\infty} \lsp e^{\lsp\gsp}  e^{-\lambda_{sd}\psi \gsp-\frac{\lambda_{sr} \psi \gsp}{(1-\rho)}}d \gsp. \nonumber 
\end{equation*}
Now after averaging over $\gsp$, the resultant expression can be found out to be:
\begin{eqnarray}
p_2=\frac{1}{1+ \frac{\psi}{\lambda_{sp}}\left(\lambda_{sd}+\frac{\lambda_{sr}}{(1-\rho)}\right)}.\label{eq:p2Final}
\end{eqnarray}
An approximate closed-form expression for $p_3$  is derived in Appendix-A. The final expression is presented in \eqref{eq:p3_Final}.
\setcounter{equation}{14}
\begin{figure*}[t!]
	\begin{IEEEeqnarray}{rcl}
		&&p_3\approx t \Bigg(\left(\frac{a}{c+\lsp}\right)^2\frac{ \lsp \lsr}{ (a+b+d \lsp)} e^{\frac{a \lsr}{c+\lsp}} \left(E_1\left(\frac{a \lsr}{c+\lsp}\right)-E_1\left(\frac{a \lsr}{c+\lsp}+a s\right)\right)+\frac{d^2 \lsp \lsr}{a+b+d \lsp} e^{s (-(a+b))-d (\lsp s+\lsr)} \nonumber\\
		&&\times(\text{Ei}(d (\lsr+\lsp s))-\text{Ei}(d \lsr+b s+d \lsp s))-\frac{\lsr (a+b)^2e^{\frac{\lsr (a+b)}{\lsp}}}{\lsp (a+b+d \lsp)}  \left(E_1\left(\frac{(a+b) \lsr}{\lsp}\right)-E_1\left(\frac{(a+b) \lsr}{\lsp}+(a+b) s\right)\right)\nonumber\\
		&&-\frac{\lsr e^{s (-(a+b))}}{\lsp s+\lsr}+\frac{\lsp \lsr e^{-a s}}{(c+\lsp) (c s+\lsp s+\lsr)}+\frac{c}{c+\lsp}\Bigg)+\frac{b^2 \lsr}{\lsp (b+d \lsp)} e^{\frac{b \lsr}{\lsp}} \left(E_1\left(\frac{b \lsr}{\lsp}\right)-E_1\left(\frac{b (\lsr+\lsp s)}{\lsp}\right)\right)\nonumber\\
		&&-\frac{d^2 \lsp \lsr}{b+d \lsp} e^{-b s-d (\lsp s+\lsr)} (\text{Ei}(d (\lsr+\lsp s))-\text{Ei}(d \lsr+b s+d \lsp s))+\frac{\lsr e^{-b s}}{\lsp s+\lsr}+\frac{c \lsp s^2}{(\lsp s+\lsr) (c s+\lsp s+\lsr)}\nonumber\\
		&&-\frac{\lsp \lsr}{(c+\lsp) (c s+\lsp s+\lsr)}-\frac{c}{c+\lsp}\,,
		\label{eq:p3_Final}\\
		&& {\normalsize\text{where $a=\frac{\lambda_{rp}}{\beta}$, $b=\frac{\psi  \lambda_{rd} }{\beta}$, $c={\psi\lambda_{sd}}$, $d=\frac{\lambda_{rd}}{\beta \lambda_{sd}}$, and $s=\frac{(1-\rho)}{\psi}$.\label{eq:param}}}
	\end{IEEEeqnarray}
	\hrulefill
	\vspace*{-0.5cm}
\end{figure*}
\setcounter{equation}{16}
Resultant expression of $\tau$ can be found out by substituting for $p_1$, $p_2$, $p_3$ and $q_2$ into \eqref{eq:tauIn}.\\
\subsection{Value of EH-factor for optimal throughput}
Throughput $\tau$ is small for $\rho=0$ (no energy harvested at the relay) and $\rho=1$ (no decoding is possible at the relay). In both these cases, the relayed  signal is not available. When $0<\rho<1$, the throughput is larger since the relayed signal is not always in outage. It is clear that the 
following optimal value of EH parameter ($\rho^{*}$) that maximizes throughput is of interest:
\begin{eqnarray}
\rho^* = \arg\displaystyle\max_{\rho} \quad \tau \label{eq:rhoOptimum}.
\end{eqnarray}
It is difficult to find an exact solution for $\rho^*$ since the lengthy expression for $\tau$ contains several nonlinear functions.
\par To obtain an expression for $\tau$ in a simplified form (say $\tau_{sim}$), we use the high-SNR approximation $\left({I\lsp}/{N_o}\gg\gamma_{th}\right)$ and also neglect the R-P link\footnote{We note that the R-P link is ignored only for the simplified analysis to obtain insights into the optimum power-splitting parameter. The relay needs to apply power control as in any other underlay system. We note that all computer simulations are performed with the interference channel from relay to primary receiver.}. We show through simulations in Fig. \ref{fig:TauVsRho}  of  Section~\ref{sec:simulations}  that throughput of a practical system that imposes the peak interference constraint at the relay is indistinguishable from one that neglects it. In other words, SN performance does not depend (or at best very loosely depends) on the statistical parameter ($\lrp$) of the R-P channel for most practical range of parameters. Intuitively, this is because of the fact that the harvested energy is very small (less than $I/\grp$) with very high probability.
\par From the above discussion, throughput can be represented in most simplified form as given in \eqref{eq:TauSimplified} (please refer the Appendix-B for derivation).
\begin{figure*}[t!]
\begin{eqnarray}
\tau_{sim}\approx0.5R_s\left(1-\left(\frac{1}{\left(1+\frac{\lsp}{\lsd\psi}\right)\left(1+\frac{\eta \lsp \rho}{\psi \lrd \lsr }\right)}+\frac{1}{\frac{\lrd  \lsp (1-\rho)}{\eta \rho  \lsr \psi}}+\frac{1}{\frac{\psi \lsd }{ \lsp}+1}\right)\right) + R_s q_2\label{eq:TauSimplified}
\end{eqnarray}
\hrulefill
\vspace*{-0.4cm}
\end{figure*}
 We omit proof of concavity due to space constraints. Solving 
$\frac{d\tau_{sim}}{d\rho}=0$.
 results in a quadratic equation which has two roots, out of which the one   between $0$ to $1$ is  given by\footnote{Since  $\lsr\ll\lsd$, it can be shown that $\rhoˆ{*}\in [0,1]$.}:
\begin{equation}
\rho^* \approx \frac{1-\sqrt{\left(1+\frac{\lsp}{\lsd\psi}\right)}\frac{ \psi \lsr\,}{ \lsp}}{1+\sqrt{\left(1+\frac{\lsp}{\lsd\psi}\right)}\frac{ \eta}{\lrd}}.\label{eq:OptRho}
\end{equation}
 \par As a special case, value of $\rho$ which maximizes the throughput if the S-D link ignored \footnote{In this case, there is no involvement of direct path and a case of two-hop transmission between nodes S to D via R.} (i.e $\lambda_{sd}\rightarrow\infty$) becomes:
\begin{eqnarray}
\rho^*_{nd} \approx \lim\limits_{d_{sd}\rightarrow\infty} \rho^* =  \frac{1-\frac{ \psi \lsr \,}{ \lsp}}{1+\frac{ \eta}{\lrd}},\label{eq:OptRhoWSD}
\end{eqnarray}
 where subscript $nd$ is used to emphasize the fact that  no direct path is present. In this case, the throughput is derived from \eqref{eq:tauIn} by using $\lsd\rightarrow\infty$, and given by:
\begin{eqnarray}
\tau_{nd} = \lim\limits_{\lsd\rightarrow\infty} \, \tau.\label{eq:TauWSD}
\end{eqnarray}
 \section{Simulation Results}\label{sec:simulations}\
In this section, we validate the derived expressions by computer simulations.  The normalized  S-R,  R-D, and  S-D distances are assumed to be $1.2$, $1.8$ and $3$ respectively. The normalized S-P and R-P distances are assumed to be $3$. Path loss exponent $\epsilon$ is assumed to be $4$. We assume $\eta = 0.7$ and $I/N_o=6$ dB unless stated otherwise.
\begin{figure}[h]
	\vspace*{-0.3cm}
	\centering
	\includegraphics[width=0.5\textwidth]{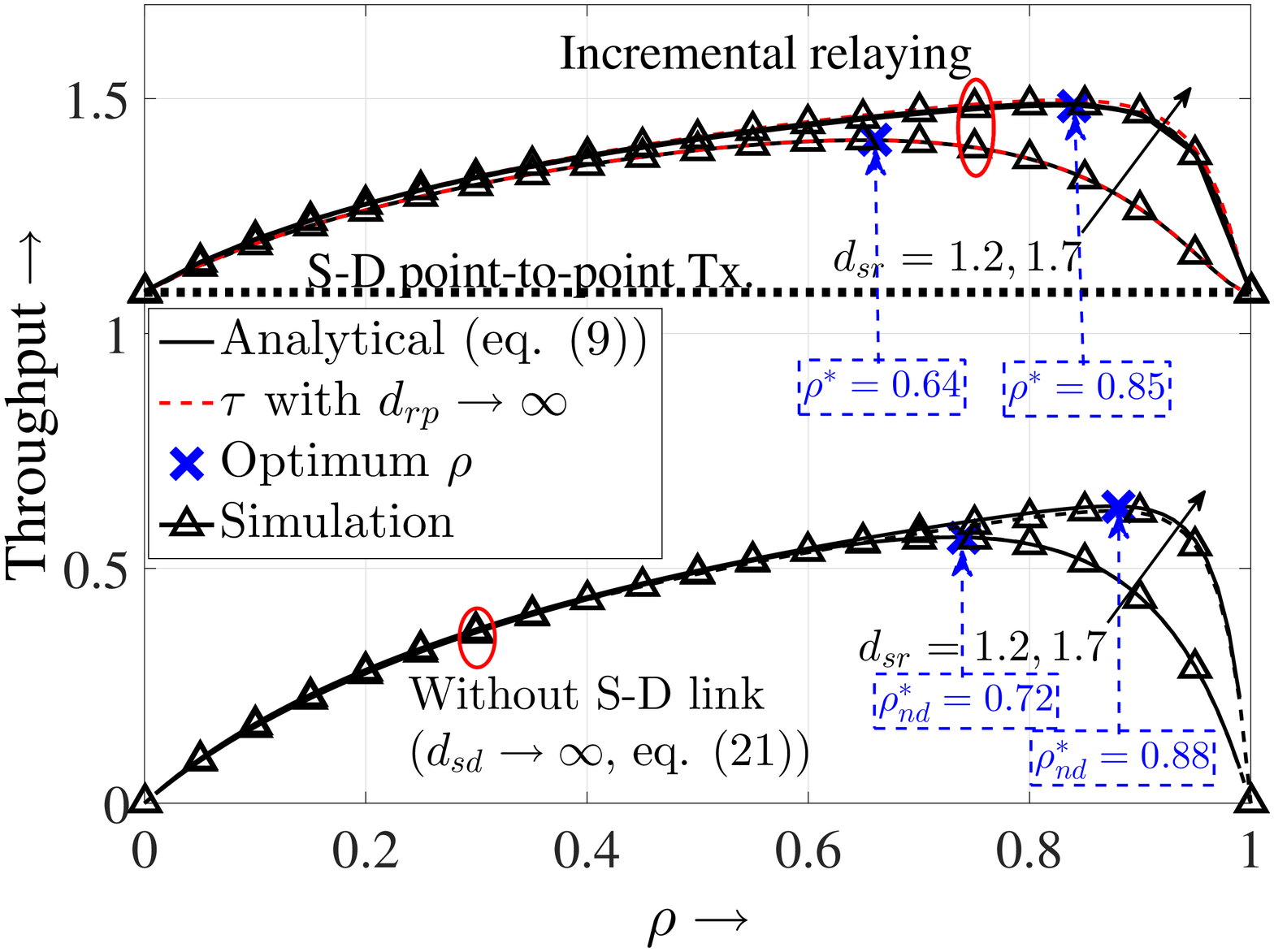}
	\caption{Throughput vs. $\rho$ for different $d_{sr}$}
	\label{fig:TauVsRho}
	\vspace*{-0.25cm}
\end{figure} 
\par The importance of optimizing $\rho$ for maximizing throughput is clearly brought out in Fig.\ref{fig:TauVsRho} (concavity is clear from the plots). The graph depicts a plot of $\tau$ versus $\rho$ for $R_s=3$ bpcu for different  $d_{sr}$. 
 \begin{figure}[h]
	\vspace*{-0.3cm}
	\centering
	\includegraphics[width=0.5\textwidth]{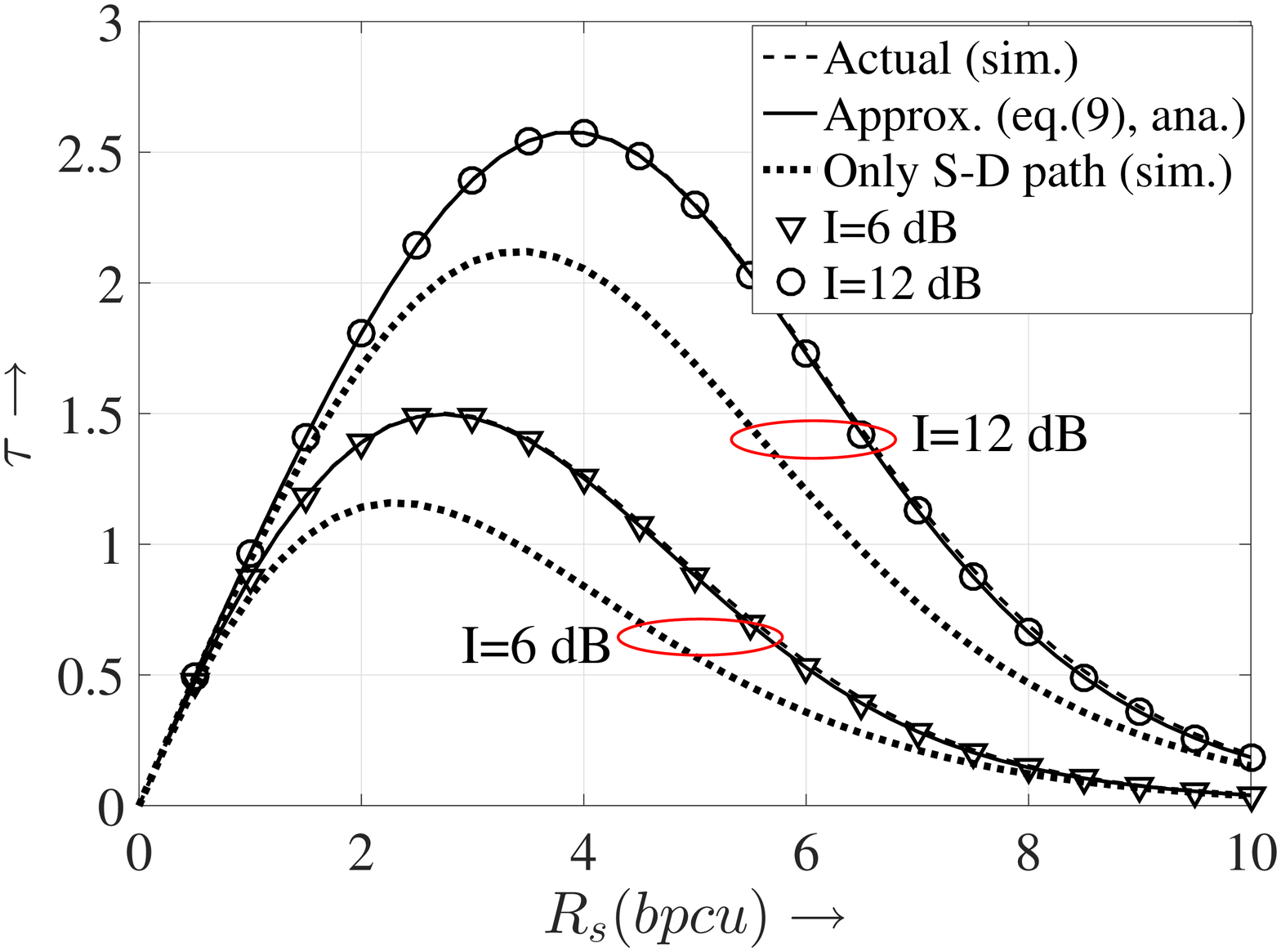}
	\caption{Throughput vs. $R_s$ for different I}
	\label{fig:TauVsRsConf}
	\vspace*{-0.5cm}
\end{figure}
\par The value of $\rho^*$ indicated by \eqref{eq:OptRho}  for $d_{sr}$  of $1.2$, and $1.7$ are $0.87$ and $0.62$ respectively, which are in close agreement with simulations.  Similarly, in the absence of the direct link,   $\rho^*_{nd}$ of  $0.89$ and $0.68$  are indicated by (\ref{eq:OptRhoWSD}), which are in close agreement with simulations. We note that $\rho^*_{nd}>\rho^*$, as can be intuitively expected.  It is clear that a) incremental relaying results in higher throughput than relay-less signalling from S to D,  b) two-hop relaying that ignores the  S-D link results in throughput that is quite poor as compared to direct S-D signalling without the relay, and c) a smaller $d_{sr}$ results in larger throughput, especially for large $\rho$ (note that $d_{sr}+d_{rd}=d_{sd}$ so that a smaller $d_{sr}$ implies a larger $d_{rd}$).
\par In Fig.\ref{fig:TauVsRsConf}, throughput is plotted versus $R_s$  for two different values of $I$.  For each point, the optimum value of $\rho^{*}$ is computed and used.  The superiority of incremental relaying over direct point-to-point transmission is apparent. Moreover, the gap between the two is higher for larger value of $I$. This happens because relayed signalling has a higher chance of non-outage for larger value of $I$.
 It be observed that an optimum value of $R_s$ exists which maximizes throughput. It is apparent from ~\eqref{eq:tauIn}, that throughput is limited by  $R_s$ when it is small, and by outage when $R_s$ is large. The optimum value needs to be obtained by numerical search.
\section{Conclusion} \label{sec:conclusion}
In this paper, we derived a closed-form expression for the throughput performance of an underlay two-hop network with a power-splitting based energy harvesting relay. We present a closed-form expression for the throughput maximizing power-splitting parameter. 
\vspace{-0.5cm}
\section*{Appendix}
\vspace{-0.5cm}
	\setcounter{equation}{25}
\begin{figure*}[t!]
	\begin{IEEEeqnarray}{rcl}
		p_3\Big|_{\hsr} &=& \frac{\lsp}{\lrd-\beta \hsr \lsd} \Big(\frac{\lrd \lsd \psi}{\lsd \lsp \psi+\lsp^2}-\beta \hsr \lsd \Big(\frac{1}{\lsp}-\beta \hsr \Big(t \Big(\frac{1}{\beta \hsr (\lsd \psi+\lsp)+\lrp}\nonumber\\
		&&-\frac{1}{\beta \hsr \lsp+\lrd \psi+\lrp}\Big)+\frac{1}{\beta \hsr \lsp+\lrd \psi}\Big)\Big)\Big)\label{eq:p3_hsr}
	\end{IEEEeqnarray}
	\vspace*{-0.6cm}
	\hrulefill
\end{figure*} 
\setcounter{equation}{27}
\begin{figure*}[t!]
	{
	\begin{IEEEeqnarray}{rcl}
	p^{d_{rp}\rightarrow\infty}_{3}&=&\frac{\left({\lsr \psi}/{\beta}\right)^2}{\lsp \left(\frac{\lrd \lsp}{\beta \lsd}+\frac{\lrd \psi}{\beta}\right)} \left(\lsr e^{\frac{\lrd \lsr \psi}{\beta \lsp}} \left(E_1\left(\frac{\lrd \lsr \psi}{\beta \lsp}\right)-E_1\left(\frac{\lrd \psi}{\beta \lsp} \left(\lsr+\frac{\lsp (1-\rho)}{\psi}\right)\right)\right)\right)\label{eq:p3_WithoutLRP}\\
	&&-\frac{\lsr \psi \left(1-e^{-\frac{\lrd (1-\rho)}{\beta}}\right)}{\lsp (1-\rho)+\lsr \psi}	-\frac{\lsp \lsr}{\frac{\lrd \lsp}{\beta \lsd}+\frac{\lrd \psi}{\beta}} \left(\frac{\lrd}{\beta \lsd}\right)^2 e^{-\frac{\lrd}{\beta \lsd} \left(\frac{\lsp (1-\rho)}{\psi}+\lsr\right)-\frac{\lrd (1-\rho)}{\beta}}\nonumber\\
	&&\times \left(\text{Ei}\left(\frac{\lrd}{\beta \lsd} \left(\lsr+\frac{\lsp (1-\rho)}{\psi}\right)\right)-\text{Ei}\left(\frac{\lrd}{\beta \lsd} \left(\lsr+\frac{\lsp (1-\rho)}{\psi}\right)+\frac{\lrd (1-\rho)}{\beta}\right)\right)\nonumber
	\end{IEEEeqnarray}}
	\hrulefill
	\vspace*{-0.4cm}
\end{figure*}
\setcounter{equation}{29}
	\begin{figure*}[t!]
	\begin{eqnarray}
	\tau_{sim} = 0.5R_s\left(1-\left(\frac{1}{\left(1+\frac{\lsp}{\lsd\psi}\right)\left(1+\frac{\eta  \lsp \rho}{\psi \lrd \lsr }\right)}+\frac{1}{\left(\frac{\lrd}{\eta \rho}+1\right) \left(\frac{ \lsp (1-\rho)}{\psi \lsr }+1\right)}+\frac{1}{\frac{\psi}{ \lsp} \left(\lsd+\frac{\lsr}{1-\rho}\right)+1}\right)\right)\label{eq:TauApproxApp1}
	\end{eqnarray}
	\hrulefill
	\vspace*{-0.4cm}
\end{figure*}
\setcounter{equation}{21}
\subsection{Derivation of $p_3$}\label{eq:p3Derivation}
In this Appendix, we derive an expression for $p_3$.
To this end, we first define $X$ as:
\begin{eqnarray}
X\triangleq \min\left(^hP, \frac{I}{\grp}\right)\hrd \label{eq:X}
\end{eqnarray}
Its Cumulative distribution function (CDF) conditioned on $^hP$ can be derived as:
\begin{equation}
\hspace{-0.04in}F_{X\big|_{^hP}}\hspace*{-0.3cm}(x)=1-e^{\frac{\lambda_{rd}x}{\beta\,P_s\,\hsr}}\Big(1-e^{-\frac{\lambda_{rp}I}{\beta\,P_s\,\hsr}}\Big(\frac{\frac{\lambda_{rd} x}{I\lambda_{rp}}}{1+\frac{\lambda_{rd}x}{I\lambda_{rp}}}\Big)\Big)\label{eq:CDFofX}
\end{equation}
From \eqref{eq:q}, $p_3$ can be expressed as:
	{\small\begin{IEEEeqnarray*}{rcl}
	p_3=\Pr\Big(\frac{P_s \hsd}{N_o}+\frac{X\big|_{P_s\hsr}}{N_o}\leq \gamma_{th},\frac{(1-\rho) P_s \hsr}{N_o} > \gamma_{th}\Big).\nonumber
	\end{IEEEeqnarray*}}
We derive $p_3$ by successive averaging over each random variable and keeping other r.vs. in terms of condition. We first average w.r.t. $X$ by using \eqref{eq:CDFofX} to get: 
\begin{IEEEeqnarray*}{rcl}
	\hspace{-0.1in}p_3&=&\E_{\C_1  }\left[ 	F_{X\big|{^hP}}\left(\gamma_{th}-\frac{P_s \hsd}{N_o}\right)\right]\nonumber\\
	&=&\E_{\C_1  }\Big[1-e^{-\frac{\lambda_{rd}}{\beta\,P_s\,\hsr}\left(\gamma_{th}-\frac{P_s \hsd}{N_o}\right)}\Big(1-e^{-\frac{\lambda_{rp}I}{\beta\,P_s\,\hsr}}\nonumber\\
	&&\hspace{0.75in}\times\Big(1-\frac{1}{1+\frac{\lambda_{rd}}{I\lambda_{rp}}\left(\gamma_{th}-\frac{P_s \hsd}{N_o}\right)}\Big)\Big)\Big]\,,
\end{IEEEeqnarray*}
where the condition $\C_1=\{\frac{(1-\rho) P_s \hsr}{N_o}>\gth,\, \frac{P_s \hsd}{N_o}\leq\gth\}$.
Unfortunately, averaging the above with respect to random variables $\hsr$ and $\gsp$ results in intractable expressions.  We need to make use of the following fact:  $1\gg\frac{\lambda_{rd}}{I\lambda_{rp}}\left(\gamma_{th}N_o-P_s \hsd\right)$ (since $P_s \hsd/N_o <\gth$ and $I\lrp\gg\lrd$).  Hence for tractability,  $\frac{\lambda_{rd}}{I\lambda_{rp}}\left(\gamma_{th}N_o-P_s \hsd\right)$ can be replaced by its mean i.e. $\frac{\lambda_{rd}}{I\lambda_{rp}}\left(\gamma_{th}N_o-\E_{P_s\hsd\leq \gth}(P_s \hsd)\right)$. $\E_{P_s\hsd\leq \gth}(P_s \hsd)$ can be derived as:
{\small\begin{equation}
	\E_{\C_2}(P_s \hsd)=\frac{I \lambda_{sp}}{\lambda_{sd}} \Big(\log \Big(\frac{\gamma_{th} \lambda_{sd} N_o}{I \lambda_{sp}}+1\Big)+\frac{\frac{\gamma_{th} \lambda_{sd} N_o}{I \lambda_{sp}}}{\frac{\gamma_{th} \lambda_{sd} N_o}{I \lambda_{sp}}+1}\Big),\label{eq:EC2}
\end{equation}}
where $\C_2=\{P_s\hsd\leq N_o \gth\}$.
Now, $p_3$ can further  represented in approximated form as: 
\begin{IEEEeqnarray}{rcl}
	p_3&\approx&\E_{\C_1}\Big[1+t e^{-\frac{\lrd (\gth N_o-\hsd P_s)}{\beta \hsr P_s}-\frac{I \lrp}{\beta \hsr P_s}}\nonumber\\
	&&\hspace*{2cm}-e^{-\frac{\lrd (\gth N_o-\hsd P_s)}{\beta \hsr P_s}}\Big], \label{eq:p3hSR}
\end{IEEEeqnarray}
where $t=1-1/({1+\frac{\lambda_{rd}}{I\lambda_{rp}}\left(\gamma_{th} N_o-\E_{\C_2}\left[P_s\hsd\right]\right)})$. The above approximation is tight for $\lrd<<I\lrp$. This is true for general system settings in underlay-CRN.  
\par Now averaging the above expression over $|h_{sr}|^{2}$ and using the expression for $P_s$ in \eqref{eq:PsOnlyInterference}, results in the following expression for $p_3$: 
\begin{IEEEeqnarray*}{rcl}
	&&p_3\approx\E_{\C_3}\Big[\frac{\lrd \left(1-e^{-\frac{\gth \lsd N_o}{P_s}}\right)}{\lrd-\beta \hsr \lsd}-\frac{(\beta \hsr \lsd)}{\lrd-\beta \hsr \lsd}\Big(1+\nonumber \\
	&&t e^{-\frac{\gth \lrd N_o+I \lrp}{\beta \hsr P_s}}-t e^{-\frac{\beta \gth \hsr \lsd N_o+I \lrp}{\beta \hsr P_s}}-e^{-\frac{\gth \lrd N_o}{\beta \hsr P_s}}\Big)\Big]\,,\nonumber 
\end{IEEEeqnarray*}
where condition $\C_3=\{\frac{(1-\rho) \hsr}{\psi}>\gsp\}$. Now let $\psi = \frac{\gth}{I/N_o}$ , $p_3$ becomes:
\begin{IEEEeqnarray*}{rcl}
	p_3\approx&\E_{\C_3}&\Big[\frac{\lrd \left(1-e^{-{\gsp\gth \lsd \psi }}\right)}{\lrd-\beta \hsr \lsd}-\frac{(\beta \hsr \lsd)}{\lrd-\beta \hsr \lsd}\Big(1\nonumber \\
	&&+t e^{-\frac{\gsp}{\beta \hsr}(\lrd\psi+\lrp)}-t e^{-\gsp\left(\psi\lsd+\frac{\lrp}{\beta\hsr}\right)}\nonumber\\
	&&-e^{-\frac{\gsp \lrd \psi}{\beta \hsr }}\Big)\Big]\nonumber
\end{IEEEeqnarray*}
\par Now averaging the above equation over $\gsp$ and then using some straightforward manipulations, the resultant expression is presented in \eqref{eq:p3_hsr}. From \eqref{eq:p3_hsr},  $p_3$ is now expressed as:
\setcounter{equation}{26}
\begin{equation}
p_3 = \int_{0}^{\infty} p_3\big|_{\hsr} \lsr e^{-\lsr \hsr} d\hsr
\end{equation}
The above integral can be simplified using the integral presented in \cite[5.1.1]{Abramowitz1964}:
 {\begin{IEEEeqnarray*}{rcl}
		\int_r^s \frac{e^{-p x}}{q x+1} \, dx = \frac{e^{\frac{p}{q}}}{q} \left(E_1\left(\frac{(q r +1)p}{q}\right)-E_1\left(\frac{(q s +1)p}{q}\right)\right)\nonumber.
	\end{IEEEeqnarray*}}
We omit the manipulations due to space limitations and present the approximated $p_3$ as in \eqref{eq:p3_Final} (top of page-4).
\subsection{Derivation of $\tau_{sim}$} \label{AppB:ApproxP3}
\par In the expression for $p_3$ in \eqref{eq:p3_Final}, $t=1-1/({1+\frac{\lambda_{rd}}{I\lambda_{rp}}\left(\gamma_{th} N_o-\E_{\C_2}\left[P_s\hsd\right]\right)})$ as defined in \eqref{eq:p3hSR}, with $\E_{\C_2}\left[P_s\hsd\right]$ given by \eqref{eq:EC2}. Clearly, $t=0$ when $\lrp\rightarrow\infty$, which enables us to simplify $p_3$. Resultant expression is given in \eqref{eq:p3_WithoutLRP} (second equation on the top of the next page).
\par Further simplification is possible when   $\frac{I\lsp}{N_o}\gg\gth$, which is the commonly encountered situation.  In this case, the arguments  of $e^{-x}\text{Ei}(x)$ and $e^{x} E_1(x)$ terms in $p_3$ increase and decrease respectively. Using  the fact that $e^{-x}\text{Ei}(x)\rightarrow 0$ for $x\rightarrow\infty$ (here $x\propto\frac{\lsp}{\psi}$), the term associated with $\text{Ei}(x)$ vanishes from the expression for $p_3$. We further use the fact that  $E_1(x)\gg E_1((x+A))$, for $A>0$ when $A$ is  very large (as it is in this case since $(1-\rho) I\lsp/(\gth N_o)>0$) . The term with $e^x E_1((x+A))$ can be neglected in the  high SNR region. From~\eqref{eq:p3_WithoutLRP}, the resultant approximated expression can be written as:
\setcounter{equation}{28}
{\small\begin{align}
	p^{d_{rp}\rightarrow\infty}_{3}\overset{A\gg0}{\approx}&\left(\frac{\psi  \lrd }{\beta \lsp}\right)^2\frac{(\lsp \lsr )}{\frac{\psi  \lrd }{\beta}+\frac{\lrd \lsp}{\beta \lsd}}e^{\frac{\psi  \lrd \lsr  }{\beta  \lsp}} E_1\left(\frac{\psi  \lrd \lsr  }{\beta  \lsp}\right)\nonumber\\
	&+\frac{\lsr }{\frac{\lsp (1-\rho)}{\psi }+\lsr } \left( e^{-\frac{ \lrd (1-\rho)}{\beta}}-1\right).	\label{eq:p3_HSNR} 
\end{align}}
By using the following very tight lower and upper bounds $e^{x}E_1(x)\geq ({1+x})^{-1}$ and 
	$e^{-{\lrd (1-\rho)}/{\rho}}\leq({1+{\lrd (1-\rho)}/{\rho}})^{-1}$
respectively into \eqref{eq:p3_HSNR}, and from \eqref{eq:p1Final},  \eqref{eq:p2Final} and \eqref{eq:q2}, $\tau$ can be approximated as $\tau_{sim}$ in \eqref{eq:TauApproxApp1}.
\par It is worth noting that the above approximation is valid for $\frac{I\lsp}{N_o}\gg\gth$. However, it continues to follow the throughput $\tau$ in other cases. To get a closed form expression, we further approximate  the above by utilizing the fact $\frac{\lrd}{\eta\rho}\gg 1-\frac{\lrd}{\eta}$ (since $\eta, \rho\leq 1$) and  $\frac{I \lsp (1-\rho)}{\gth N_o\lsr}\gg1$ except when $\rho\approx1$ and represented in \eqref{eq:TauSimplified} (top of the page-4). Please note that $\rho\approx1$ is unlikely, since it results in outage of the relayed link.  
\vspace*{-0.15cm}
\bibliographystyle{ieeetr}
\bibliography{IEEEabrv,MRC_bib}
\end{document}